
\magnification=1220
\footline={\ifnum\pageno=1\hfill\else\hfill\rm\folio\hfill\fi}
\font\title=cmr10 at 18pt
\baselineskip=18pt
\rightline{DFTUZ 94/03}
\rightline{LPTHE Orsay-94/12}
\rightline{January 94}
\def\dsl{\raise.15ex\hbox{/}\kern-.57em\partial}
\def\Dsl{\,\raise.15ex\hbox{/}\mkern-13.5mu D} 
\def\Asl{\,\raise.15ex\hbox{/}\mkern-13.5mu A} 
\def\Bsl{\,\raise.15ex\hbox{/}\mkern-13.5mu B} 
\def\la{\left\langle}
\def\ra{\right\rangle}
\def\sub#1 {${\rm \underline #1 }$}
\def\nl{ \hfill\break }
\vskip 1.0cm
\centerline{{\title PHASE DIAGRAM OF AN}}
\vskip 0.1cm
\centerline{{\title SU(2) x SU(2) SCALAR-FERMION MODEL}}
\vskip 0.1cm
\centerline{{\title WITH MASSLESS DECOUPLED DOUBLERS}}
\vskip 1.5cm
\centerline{\bf J.L. Alonso, F. Lesmes, E. Rivas}
\centerline{\it Departamento de F\'\i sica Te\'orica, Universidad de Zaragoza,}
\centerline{\it 50009 Zaragoza, Spain.}
\vskip 0.5cm
\centerline{\bf Ph. Boucaud}
\centerline{\it Laboratoire de Physique Th\'eorique et Hautes
Energies\footnote{$^*$}{Laboratoire associ\'e au CNRS}.}
\centerline{\it Universit\'e de Paris XI, 91405 Orsay Cedex, France}

\vskip 2.5cm

{\bf Abstract}.
We present the phase structure of the chiral
$SU(2) \times SU(2)$ scalar-fermion model on the lattice using
the Zaragoza proposal for chiral fermions. The numerical result
agrees with an analytic study based on the use of weak and
strong yukawa coupling expansions combined with
the mean-field approach.
The phase diagram consists of four phases:
paramagnetic (PM), ferromagnetic (FM), antiferromagnetic (AFM) and
ferrimagnetic (FI). The transition lines separating these four phases
intersect at one quadruple point.

\vfill \eject

\centerline{\bf 1. Introduction}
\vskip 0.5cm

The formulation of a Chiral Gauge Theory (CGT) on the lattice suffers
from the well-known doubling
problem  [1]. Several ways of dealing with this problem have
been reviewed in reference [2]. Among them, the Zaragoza proposal [3,4]
is a viable candidate for describing chiral fermions on the lattice.
This proposal belongs to a category of models wherein, as in the Rome
approach [5], gauge symmetry is explicity broken by the regularization
and therefore additional counterterms must be included in order to
restore the gauge invariance in the continuum limit.
The distinguishing feature of this approach is that the global chiral
symmetry  is preserved, and the doubler fermions are
free and massless. These properties
simplify the  calculation of some
important parameters of the Standard Model
(the $S$, $U$ and $\Delta\rho$ parameters)[6].

In the case of Chiral Yukawa models (CYM), this regularization method
preserves  the invariance under discrete
rotations and translations, but also all the important symmetries
present in the continuum: hermiticity and global chiral symmetry.
Hence it is particularly well suited to study those models [4].

In this paper, we investigate the phase structure of
a ${SU(2)}_L \times {SU(2)}_R$ CYM which is essentially
the fermion-scalar sector of the electroweak theory.
We decided to freeze the radial mode of the scalar field,
which  corresponds to the choice of
an infinite bare quartic coupling. Knowledge from the
pure  $\Phi^4$ theory suggests that such a model belongs to the same
universality class as the models with finite quartic coupling.
For a comparison of the phase structure
of different models see references [2,7-10].

\vskip 1.5 cm
\centerline{\bf 2.Model, symmetries  and limiting case}
\vskip 0.5cm

Our action $S(\Psi,\Phi)$ is given by
$$
S(\Psi,\Phi) = S_B(\Phi) + S_F(\Psi) + S_Y(\Psi,\Phi), \eqno (1) $$
where
$$
S_B(\Phi) = - {k \over 2} \sum_{x,\mu}{\rm Tr }\left( \Phi^+_{x+\hat\mu}\Phi_x
+ \Phi^+_x\Phi_{x+\hat\mu} \right) \eqno (2) $$
is the kinetic term for the scalar fields, $\Phi$, which are $2\times 2$
$SU(2)$ matrices.
$$
S_F(\Psi) = {}  {1 \over 2} \sum_{\Psi,x,\mu}
\left( \overline \Psi_x \gamma_\mu \Psi_{x+\hat\mu} - \overline
\Psi_{x+\hat\mu} \gamma_\mu \Psi_x\right) \eqno (3)
$$
is the fermionic kinetic term where   $\sum_\Psi$ stands
for the summation over $n_f$ doublets of Dirac fermions, and  finally,
$$
S_Y(\Psi,\Phi) = {}  y \sum_{\Psi,x} \left(
\overline \Psi^{(1)}_{Lx} \Phi_x \Psi^{(1)}_{Rx} +
\overline \Psi^{(1)}_{Rx} \Phi^+_x \Psi^{(1)}_{Lx}\right) \eqno (4)
$$
is the Yukawa interaction.
The $L$ and $R$ indices refer to the left and right components
of the fermion fields.
In this interaction term,
the way to implement the decoupling of the doubler fermions is based on the
use of the quasi-local field component
$\Psi^{(1)}$ given by
$$
\Psi^{(1)}_x =
{1 \over 2^d} \hskip -2.5mm
\sum_{x'\in\ hc\left( x \right)} \hskip -3mm
\Psi_{x'}. \eqno (5)
$$
Here $hc\left( x \right)$ is the elementary hypercube that starts from
the site $x$
in the positive direction. To understand why we have chosen $\Psi^{(1)}$
to describe the fermion interactions in Eq. (4) let us go to momentum space.
There $\Psi^{(1)}$ is given by
$$
\Psi^{(1)}(q) = F(q)\Psi(q),\qquad
F(q) = \prod_\mu f(q_\mu), \qquad
f(q_\mu) = \cos({q_\mu\over 2}),  \qquad
q \in (-\pi,\pi]^d. \eqno (6)
$$
Note that $\Psi^{(1)}$ is the original field $\Psi$ modulated by a form
factor $F$ which is responsible for the decoupling of the doublers
at the tree level in the continuum limit. This is achieved by forcing the
form factor to vanish for momenta corresponding to the doublers
(note that $f(0) = 1$ and $f(\pi) = 0$).
Other form factors are possible but eq. (6) corresponds
to the most local possible choice for the smearing in position space,
given by eq. (5).
We have proved perturbatively [3,4] that in the continuum limit only one
fermion (for each flavour) is coupled to the scalar. The doublers stay
free and massless.

The action (1) has a global chiral symmetry and is also invariant under
a shift transformation of the fermion
fields. This shift-symmetry  can be used to argue, if a continuum limit
exists,
the non-perturbative decoupling of the doublers (see Ref. 4 for more
details).

A few words are needed about the symmetries of the phase diagram
because, in this model, they are not exactly the same as in other
regularizations and have relevant consequences.

Some of the usual symmetries are present.
The action is invariant under the transformation
$\Phi_x \to -\Phi_x $, $ y \to -y$, so we can restrict ourselves to
$y\ge 0$
with no
loss of generality.
For $y = 0$ the action is invariant under $k \to -k$,\quad $\Phi_x \to
\epsilon_x\Phi_x$,\quad $\epsilon_x = {(-1)}^{x_1+x_2+x_3+x_4}$; the
scalar and fermionic fields are decoupled and
the model reduces to   a $\Phi^4$ model with the radial mode frozen:
two critical points are present at $k = \pm k_c$.

{\it But  the symmetry $k \to -k$, $ y \to -iy$, usual in other
models, is absent here} [4].
As a consequence  the phase transition line
PM--AFM is not determined by the phase transition line PM--FM,
as often happens [7].  The  simple mean field
computation that we shall present in  sect. 3 is able to detect  the crossing
of these lines  and the presence of  a ferrimagnetic phase.

Concerning the limiting case $y \to \infty$, we have a peculiarity
similar to that in the Chiral Yukawa model with hypercubic coupling [9].
A rescaling of the fermionic field
$ \Psi \to \left(1/\sqrt y\right)\Psi$, suppresses, as always, the fermion
kinetic term but, in our case, the fermionic fields on different sites are
still coupled
by the Yukawa interaction $S_Y$  (because of the smearing in $\Psi^{(1)}$).
The fermion can  propagate and  does not decouple from the scalar field when
$y$ goes to $  \infty$. In this limit the model is
not equivalent to a pure $\Phi^4$ model.

\vskip 1.5 cm
\centerline{\bf 3. Yukawa coupling expansions and the mean field technique}
\vskip 0.5cm

To evaluate the free energy, we have used conventional mean-field
methods  for the scalar field  [11]
together with weak and strong coupling expansions for the Yukawa
interaction. At small $y$, the Yukawa term:
$$
{\rm e}^{-S_Y}\ =\  {\exp} \left[- y \sum_{\Psi,x} \left(
\overline \Psi^{(1)}_{Lx} \Phi_x \Psi^{(1)}_{Rx} +
\overline \Psi^{(1)}_{Rx} \Phi^+_x \Psi^{(1)}_{Lx}\right)\right], \eqno (7)
$$
is straightforwardly expanded in powers of $y$ up to the four fermion
interaction (i.e. up to $y^2$).

On the other hand, for large values of $y$, we first introduce auxiliary
fermionic fields
$\overline\eta_x$ and $\eta_x$ using the following identity:
$$
\exp \left\{ -y \left(
\overline \Psi^{(1)}_x \Phi_x P_R \Psi^{(1)}_x +
\overline \Psi^{(1)}_x \Phi_x^\dagger P_L \Psi^{(1)}_x \right) \right\} =
$$
$$
y^{2^{d/2}} \int d\eta_x d\overline\eta_x \exp \left\{ {1 \over y} \left(
\overline \eta_x \Phi_x^\dagger P_R \eta_x + \overline \eta_x \Phi_x P_L
\eta_x
 \right) +
\overline \Psi^{(1)}_x \eta_x + \overline \eta_x \Psi^{(1)}_x \right\},
\eqno (8)
$$
and proceed with an expansion of the exponential term in powers of
$1/y$. Here again, we have kept all the terms up to the four fermion
interaction (i.e. up to $1/y^2$).

The next step is to make the scalar field dependence of $S_B(\Phi)$ linear
by using the auxiliary fields $V_x, V_x^\dagger, A_x, A_x^\dagger$
(N is the number of sites):
$$
\exp\left\{ {k\over 2} \sum_{x,\mu} {\rm Tr }
\left( \Phi^\dagger_x \Phi_{x+\hat\mu} +
\Phi^\dagger_{x+\hat\mu} \Phi_x \right) \right\} =
$$
$$
\left( {1 \over 2\pi}\right)^{8N} \int
\left[ dV dV^\dagger dA dA^\dagger \right]
\exp\left\{ {k\over 2} \sum_{x,\mu}
{\rm Tr }\left( V^\dagger_x V_{x+\hat\mu} +
V^\dagger_{x+\hat\mu} V_x \right) \right.
$$
$$
\left. {} + {i \over 4} \sum_x {\rm Tr }\left[ A^\dagger_x (\Phi_x - V_x) +
A_x (\Phi^\dagger_x - V^\dagger_x) \right]
\right\} . \eqno (9)
$$

In both cases (for small or large $y$ ), the functional integration over
the scalar field $\phi$ can be performed and the fermion terms
exponentiated.
Before doing the integration over the fermion fields (in order to
convert the Yukawa model into a purely bosonic model), we need to decouple
the composite fields of the type $\overline\Psi_x\Psi_x$ (which appear in the
four fermion interaction) by using  identities like [12]:
$$
\exp \left\{ {1\over2}
{\left( \sum_\Psi \overline \Psi^{(1)}_x \Psi^{(1)}_x \right)}^2
M_x^{-1} \right\} =
\int {d\lambda_x \over \sqrt{2\pi}} \exp \left\{ - {1\over2} {\lambda_x}^2 +
M_x^{-1/2} \sum_\Psi \overline \Psi^{(1)}_x \Psi^{(1)}_x \lambda_x \right\}
\eqno (10) $$

As a result of the bilinear structure of the action in the fermionic
fields, it is now possible
to carry out  the fermionic integrations in the partition function.
The remaining integrals  are approximated by
the mean field saddle point method: we look for a translationally
invariant saddle point with  a constant and a staggered piece,   we then
substitute the auxiliary fields
$V_x$ and $ V^\dagger_x$ with
$v + v_{st}\epsilon_x$,
$ A_x$ and $ A^\dagger_x $ with $-i( \alpha + \alpha_{st}\epsilon_x) $ and
$ \lambda_x$ with $\lambda + \lambda_{st}\epsilon_x$.
This approach yields the free energy per unit volume,
$$
\eqalign{{\cal F}\ =\ -{1 \over N} \log Z \ =\ &{1 \over 2} (\lambda^2 +
\lambda_{st}^2) - 2 \overline k (v^2 - v_{st}^2) +  \alpha v +
\alpha_{st} v_{st} \cr
&- {1 \over 2}\left[ u(\alpha + \alpha_{st}) +
u(\alpha - \alpha_{st}) \right] - {n_f \over 2} 2^{d/2} I,} \eqno (11)
$$
with $\overline k = kd$,
$u(\beta) = \log{2 \over \beta} I_1(\beta)$, where $I_1$ is
the modified Bessel function of order 1.
The function $I$ comes from the fermionic determinant and has
different expressions for strong and weak yukawa coupling, $I=I_S$ and
$I=I_W$ respectively with:
$$
I_W  =\ \int {{\rm d}^d p\over {(2\pi)}^d}
\log\left\{ {\left[ s^2(p) + y^2 z \tilde z F^2(p) F^2_\pi(p) \right]}^2 +
y^2 {\left( {z+\tilde z \over 2} \right)}^2 s^2(p)
{\left( F^2(p) - F^2_\pi(p) \right)}^2 \right\}, \eqno (12)
$$

$$
I_S = \int {d^d p\over {(2\pi)}^d}
\log\left\{ {\left[ z \tilde z s^2(p) + y^2 F^2(p) F^2_\pi(p) \right]}^2 +
y^2 {\left( {z+\tilde z \over 2} \right)}^2 s^2(p)
{\left( F^2(p) - F^2_\pi(p) \right)}^2 \right\}. \eqno (13)
$$
where  $F_\pi(p) = F(p_1+\pi,\dots,p_d+\pi)$, and
$$
s^2(p) = \sum_\lambda \sin^2 p_\lambda,
$$
$$
z = \dot u(\alpha + \alpha_{st}) -
(\lambda + \lambda_{st}) \sqrt{ \ddot u(\alpha + \alpha_{st}) },
$$
$$
\tilde z = \dot u(\alpha - \alpha_{st}) -
(\lambda - \lambda_{st}) \sqrt{ \ddot u(\alpha - \alpha_{st}) }. \eqno (14)
$$

The saddle point equations are obtained by requiring that  the free
energy is extremal for
$v$, $\alpha$, $\lambda$, $v_{st}$, $\alpha_{st}$ and $\lambda_{st}$.
The free energy in the presence of source terms can be evaluated following
the same steps, and the mean field predictions for the order parameters
are obtained by taking derivatives with respect  to these sources. For
instance,
for a lattice with $N$ sites, with the
definitions
$$
\la\Phi\ra = \la{1\over N}\sum_x\Phi_x\ra, \qquad
\la\Phi_{st}\ra = \la{1\over N}\sum_x\epsilon_x\Phi_x\ra ,\eqno (15)
$$
we  find $\la \Phi \ra = v$, $\la \Phi_{st} \ra = v_{st}$ if we align
the symmetry breaking direction along the unity matrix.
Looking numerically for the saddle point values and the free energy
for several choices for
the parameters $\overline k$ and $y$, we have found four phases, (see
fig. 1):

\leftline{Paramagnetic (PM): Here, $\la \Phi \ra = \la \Phi_{st} \ra =
0$.}
\leftline{Ferromagnetic (FM): In this phase $\la \Phi \ra \not= 0$ and
$\la \Phi_{st} \ra = 0$.}
\leftline{Antiferromagnetic (AFM): This phase is characterized by $\la \Phi
\ra = 0$ and $\la \Phi_{st} \ra \not= 0$. }
\leftline{Ferrimagnetic (FI): Both parameters different from zero.}

The expansion of the term $e^{-S_Y}$
is essentially an expansion in $y\ \overline \Psi^{(1)}_x \Psi^{(1)}_x$
or in $1/y \ \overline\eta_x \eta_x$,
and as a consequence
we   expect that  our results are not too bad in  regions where
$ \left| y\la \overline \Psi^{(1)}_x \Psi^{(1)}_x \ra \right| < 1$
at small $y$ or where
$ \left| (1/y)\la \overline \eta^{(1)}_x \eta^{(1)}_x \ra \right| < 1$
at large $y$.
In the figure we have only shown the regions of $y$ where these
conditions  are fulfilled.

One last comment about the mass of the physical fermions in the
FM phase, $m_F$ (in lattice units). For small $y$, it can be shown, from
eqs. (11,12), that $m_F^2$ =  $y^2\ z^2$.
For large $y$, eqs.  (11,13) gives   $m_F^2$ =  $y^2/z^2$. Therefore, as
  usual [7],
we can distinguish  a  weak and a  strong  FM regions.

\vskip 1.5 cm
\centerline{\bf 4. Phase transition lines}
\vskip 0.5cm

A sampling of the values of the order parameters for several values of
$y$ suggests  that the change in the order
parameters occurs continuously
when crossing the phase transitions.
Close to the phase transitions, this allows a
linearization of the saddle point equations in the
mean field variables which
vanish at the phase transition (those are
$v$, $\alpha$, $\lambda$ for the FM--PM and AFM--FI phase transitions and
$v_{st}$, $\alpha_{st}$, $\lambda_{st}$ for the AFM--PM and FM--FI phase
transitions).
In  the phase diagram plotted in the figure,
the FM--FI and AFM--FI transition lines have
been obtained  with a numerical solution of the linearized equations.
For the FM--PM and the
PM--AFM transitions an analytical solution is possible; the transitions
are of second order and the critical lines are:
$$
\overline k_c = \pm 1  -  y^2 2^{d/2}{n_f \over 2} I_\pm, \eqno (16)
$$
where the upper signs stand
for the FM--PM transition and the lower signs for the PM-AFM.
For  $d = 4$,
$$
I_+ = \int {d^4 \theta \over {(2\pi)}^4} {F^4(\theta) \over s^2(\theta)}
\simeq 1.61 \times 10^{-2} \eqno (17)
$$
$$
I_- = \int {d^4 \theta \over {(2\pi)}^4} {F^2(\theta) F_\pi^2(\theta)
\over s^2(\theta)}
\simeq 8.4\times 10^{-5} \eqno (18)
$$

Because of the very small value of $I_-$ the PM--AFM transition line is
almost straight.
{\it One would expect this behaviour
in any theory in which the doubler fermions
are decoupled}.
In fact, in the
mean field approximation, without any form factor, in the AFM phase, the
Yukawa term couples the scalar mean field, $v_{st}$, to the fermion sector
through a coupling of the type
$ \overline\Psi_{k'}v_{st}\Psi_k\delta(k-k'-\pi)$.
Because of the factor
$\delta(k-k'-\pi)$, the only coupling between the scalar and
the fermion
is through a vertex involving simultaneously
the physical and the doubler fermions. Therefore, if
somehow the doubler fermions are
effectively decoupled from the scalar sector, {\it simultaneously the
scalar sector is decoupled from the physical fermion}, and the PM--AFM phase
transition line will be straight.

In our model it is easy to understand how this phenomenon occurs. In $I_+$,
$F^4(\theta)$ admits the contribution of the physical fermions and kills the
contribution of the doublers, whereas in $I_-$, $F^2(\theta)F_\pi^2(\theta)$
kills the contribution of the doublers but also that of the physical
fermions. The PM--AFM transition is nearly independent of $y$. We can
compare with the case of the naive fermions where $F^4(\theta)$ =
$F^2(\theta)F_\pi^2(\theta)$ = 1,
 so $I_+$ = $I_-$ and we recover the $k \to -k$, $ y \to -iy$ symmetry;
the transition lines PM-FM and PM-AFM are now parallel, the FI phase
and the quadruple point are not seen in the mean field approximation.

Note also that  the average slope of the  PM--FM transition line
in our case is smaller than when the doublers are coupled.
Consequently the value of the bare Yukawa coupling at the quadruple
point is large. Moreover, perturbative renormalisation group
arguments  are in favor of an increase of the  renormalized  coupling
constant when the number of coupled fermions decreases. This could be
relevant for the upper bounds of the physical masses.

\vskip 1.5 cm
\centerline{\bf 5. Monte-Carlo results}
\vskip 0.5cm

We have performed an exploratory numerical simulation of this model
on   lattices of size $4^4$  and $8^4$. We have used the Hybrid Monte Carlo
algorithm [13]  with two flavour doublets (an even number is required
for the algorithm).
The fermion matrices was inverted with the Conjuguate
Gradient algorithm. The length of the molecular dynamics trajectories was
selected with
a random distribution with a mean value of 10 steps.
For chosen values of $y$, we scan over $k$ in a region
where a  phase transition is predicted by the mean field results
given in the
previous section. For each simulation in the $(y,k)$ plane, the number of
configurations varies between 3000 and 14000 after  having discarded
about
1000 trajectories  for thermalisation. We have measured the order
parameters and the  associated susceptibilities.
In the  large $y$ region, we have only
checked that we have a ferrimagnetic phase when  $k$ is sufficiently
negative without determining precisely the transition point between
the FM and the FI phases.
In the small $y$ region, the measured
transition points are plotted on the figure  and compare well  with the
results from the mean field predictions given by the dashed lines.

\vskip 1.5 cm
\centerline{\bf 6. Conclusions}
\vskip 0.5cm
 We have determined the phase structure of an $SU(2)\times SU(2)$
fermion-scalar model with the globally chiral invariant model
proposed in [3,4] to decouple the doublers. The numerical and mean field
results are in good agreement. We have found four phases (PM, FM, AFM,
FI) with a quadruple point. The mean field  calculation is
 able to predict the FI phase and the quadruple point. The PM--AFM
transition line is  nearly independent of $y$, this property
is intimately related with the decoupling  of the doublers.

The global chiral symmetry  is preserved by this model
and the
decoupling mechanism allows the PM--FM transition line to reach
rather large bare yukawa couplings.
Thus, the questions of a bound on the
mass of a heavy fermion induced by  a strong Yukawa interaction and its
influence on the Higgs mass  should be fruitfully investigated along
this way.
These issues are currently
under investigation.

\vskip 2.0 truecm

{\bf Acknowledgements:}
This work was partially supported by the CICYT (proyecto AEN 90--0030).
We would like to thank A. Taranc\'on for enlightening discussions.
J.-L. Alonso and F. Lesmes wish to thank the LPTHE in ORSAY
for its kind hospitality.

\vskip 1.5 cm
\centerline{\bf References}

\item{\bf [1]} L.H. Karsten and J. Smit,
{\it Nucl. Phys.} {\bf B183} (1981) 103.\nl
H.B. Nielsen and M. Ninomiya,
{\it Nucl. Phys.} {\bf B185} (1981) 20; {\bf B193} (1981) 173.

\item{\bf [2]} D.N. Petcher,
{\it Nucl. Phys. B (Proc. Suppl.)} {\bf  30} (1993) 50.

\item{\bf [3]} J.L. Alonso, Ph. Boucaud, J.L. Cort\'es and E. Rivas,
{\it Phys. Rev.} {\bf D44} (1991) 3258;
{\it Mod. Phys.} Lett. {\bf A5} (1990) 275;
{\it Nucl. Phys. B (Proc. Suppl.)} {\bf 17} (1990) 461.

\item{\bf [4]} For a comprehensive review of this proposal see:
J.L. Alonso, Ph. Boucaud, J.L. Cort\'es, F. Lesmes and E. Rivas,
{\it Nucl. Phys. B (Proc. Suppl.)} {\bf 29B,C} (1992) 171.

\item{\bf [5]} A. Borelli, L. Maiani, G.C. Rossi, R. Sisto and M. Testa,
{\it Nucl. Phys.} {\bf B333} (1990) 335.\nl
 L. Maiani, G.C. Rossi and M. Testa,
{\it Phys. Lett.} {\bf B292} (1992) 397.

\item{\bf [6]} J.L. Alonso, Ph. Boucaud, J.L. Cort\'es, F. Lesmes
and E. Rivas, {\it Nucl. Phys.} {\bf B407} (1993) 373.

\item{\bf [7]} W. Bock, A.K. De, K. Jansen, J. Jersak, T. Neuhaus
and J. Smit,
{\it Nucl. Phys.} {\bf B344} (1990) 207.\nl
J. Shigemitsu,
{\it Nucl. Phys. B (Proc. Suppl.)} {\bf 20} (1991) 515.\nl
Toru Ebihara and Kei-ichi Kondo,
{\it Prog. Theor. Phys.} {\bf 87} (1992) 1019.\nl
C. Frick, T. Trappenberg, L. Lin, G. M\"unster, M. Plagge, I. Montvay
and H. Wittig, preprint DESY 92-11.

\item{\bf [8]}
W. Bock, C. Frick, J. Smit and J.C. Vink,
{ \it Nucl. Phys.} {\bf B400} (1993) 309.\nl
W. Bock, J. Smit and J.C. Vink,
{\it Phys. Lett.} {\bf B291} (1992) 297.\nl
J. Smit, {\it Nucl. Phys. B (Proc. Suppl.)} {\bf 29B,C} (1992) 83.

\item{\bf [9]} I--H. Lee, J. Shigemitsu and R.E. Shrock,
{\it Nucl. Phys.} {\bf B330} (1990) 225.

\item{\bf [10]} S. Zenkin, KYUSHU-HET-8, September 1993.

\item{\bf [11]} J.M. Drouffe and J.B. Zuber,
{\it Phys. Rep.} {\bf 102} (1983) 1.

\item{\bf [12]} H. Kluberg--Stern, A. Morel and B. Petersson,
{\it Nucl. Phys.} {\bf B215 [FS7]} (1983) 527.

\item{\bf [13]}S. Duane, A.D. Kennedy,  B.J. Pendleton and D. Roweth,
{\it Phys. Lett.} {\bf B195 } (1987) 216.

\vskip 1.5 cm
\centerline{\bf Figure Caption}

\item{Fig.1} Phase diagram for the  ${SU(2)}_L \times {SU(2)}_R$
Chiral Yukawa model, for $n_f = 2$ doublets of Zaragoza fermions.
The dashed lines are the mean field results for the transition lines.
The open symbols are the transition points determined by the Monte Carlo
simulation.
The square correspond to the FM--PM or AFM--FI transitions, the circles
to the AF--FM or FI--FM ones. Because of CPU limitation, we have not
succeeded  in  a precise determination of the position  of the quadruple
point.
\end